\documentclass[graybox]{svmult}
\usepackage{type1cm}
\usepackage{makeidx}
\usepackage{graphicx}
\usepackage{multicol}
\usepackage[bottom]{footmisc}
\usepackage{newtxtext}
\usepackage{newtxmath}

\makeindex

\begin{document}

\title*{Statistical and dynamical bimodality in multifragmentation reactions}
\author{S. Mallik, G. Chaudhuri, F. Gulminelli, S. Das Gupta}
\institute{S. Mallik \at Variable Energy Cyclotron Centre, 1/AF Bidhan Nagar, Kolkata 700064, India\\
\email{swagato@vecc.gov.in}
\and G. Chaudhuri \at Variable Energy Cyclotron Centre, 1/AF Bidhan Nagar, Kolkata 700064, India
\and F. Gulminelli \at LPC Caen IN2P3-CNRS/EnsiCaen et Universite, Caen, France
\and S. Das Gupta \at Physics Department, McGill University, Montr{\'e}al, Canada H3A 2T8}

\maketitle
\abstract{: The bimodal behavior of the order parameter is studied in the framework of Boltzmann-Uehling-Uhlenbeck (BUU) transport model. In order to do that, simplified yet accurate method of BUU model is used which allow calculation of fluctuations in systems much larger than what was considered feasible in a well-known and already existing model. It is observed that depending on the projectile energy and centrality of the reaction, both entrance channel and exit channel effects can be at the origin of the experimentally observed bimodal behavior. Both dynamical and statistical bimodality mechanisms are associated in the theoretical model to different time scales of the reaction, and to different energy regimes.}
\section{Introduction}
The bimodal behaviour of the order parameter is an important signature of first order phase transition \cite{gross,Gulminelli_Phase_transition,Dasgupta_Phase_transition}. Phase transitions occur in very large systems, but in nuclear physics, practical theoretical calculations (and heavy ion reaction experiments) need to be done with finite systems.\\
\indent
The largest cluster is an important order parameter for studying nuclear liquid gas phase transition in intermediate energy heavy ion reactions. For an infinite system, the normalized order parameter is $1$ in the liquid phase and it suddenly drops to $0$ when the system crosses the transition temperature. From, statistical point of view, bimodality is a result of the singularity (infinite system) being replaced by the smearing (finite system). At low temperature, the largest cluster probability peaks at the liquid side where as at high temperature it is limited to the gas side only. But in a small range of intermediate temperature one can expect a a double humped distribution (hence the name bimodality). Different theoretical studies (from lattice gas model and statistical models) \cite{zeroes,Gulminelli_bimo,Chomaz_bimo,bimo_add,Pleimling,lehaut10,bimodal_gargi}as well as experimental observations \cite{Bruno,Bonnet,Borderie} confirm this kind of signature of thermal phase transition in heavy ion reactions.\\
\indent
On the other side, some other theoretical calculations \cite{zbiri,lefevre,lefevre2,Mallik16} as well as experimental measurements \cite{Pichon1,Lautesse,Lopez} conclude that, a memory of the entrance channel is clearly present and thermal equilibrium is not achieved. The signal was interpreted in these studies as a dynamical bifurcation of reaction mechanism, induced by fluctuations of the collision rate, which leads to fluctuations of the collective momentum distribution as expected in complex nonlinear dynamical systems.\\
\indent
Therefore, the origin of the experimentally observed bimodality is still not clear completely. In the previous dynamical approaches used \cite{zbiri,lefevre,lefevre2,Mallik16} to study the
bimodality phenomenon, the collision final state was determined by the semiclassical one-body transport equation itself, considering simulations evolving until asymptotic times. However, these approaches lack the necessary correlations to properly treat fragment formation in the exit channel, even if they are known to very well describe the entrance channel of heavy-ion reactions at intermediate energy. For this reason, to have a quantitative reproduction of experimental data, the secondary decay of the dynamically formed primary fragments
is typically treated in two-steps calculations, coupling the transport dynamics to a statistical model (or "afterburner"). As the primary interest is phase transition in nuclear matter due to the nuclear force alone, most theoretical models have considered symmetric nuclear matter where the Coulomb force is switched off \cite{Dasgupta,Bugaev}.  Here we follow the same practice.\\
\indent
In order to study the dynamical stage for phase transition, one needs to simulate collisions between fairly large nuclei. In order to do that, a simplified yet accurate method of transport model based on Boltzmann-Uehling-Uhlenbeck (BUU) equation is developed recently \cite{Mallik10} which allows calculation of fluctuations in systems much larger than what was considered feasible in a well-known and already existing model \cite{Bauer}. For studying de-excitation phase, Canonical Thermodynamical Model (CTM) is used.\\
\indent
From this theoretical study it is observed that, depending on the incident energy and impact parameter of the reaction, dynamical as well as statistical bimodality mechanisms can appear, meaning that the different scenario proposed in the literature are both potentially observable in heavy ion data. Specifically, fluctuations in the stopping dynamics in central collisions lead to different reaction mechanisms that can
coexist in the the sample characterized by a well-defined value of the impact parameter. This gives rise to a bimodal behavior
of the largest cluster probability distribution that can survive to the secondary de-excitation if the deposited energy is low enough, which
happens at incident energies around the Fermi energy domain (40 MeV/nucleon). At higher incident energies (100 MeV/nucleon), focusing on
binary mid-peripheral reactions, the fluctuations in the energy deposition leads to an excitation energy distribution for the
quasispectator source which is close to the liquid gas phase transition range. For these events, local equilibrium is achieved and a
thermal bimodality is observed in agreement with statistical expectations.\\
\indent
This article is structured as follows. The modifications in the fluctuation included BUU model which is essential for studying liquid gas phase transition is introduced in section 2, the coupling conditions between the dynamical and statistical treatment are explained in section 3, the results concerning the different conditions of dynamical and statistical bimodal behavior are described in section 4 and 5 respectively, finally summary is presented in section 6.
\section{Improvement in BUU model with fluctuation}
The BUU transport model calculation \cite{Mallik10,Dasgupta_BUU1,Mallik_world_scientific,Mallik11} for heavy ion collisions starts with two nuclei in their respective ground states approaching each other with specified velocities and impact parameters. The ground state energies and densities of the projectile (mass number $A_p$) and target (mass number $A_t$) nuclei are constructed using the Thomas-Fermi approximation \cite{Mallik_world_scientific}. The Thomas-Fermi phase space distribution is then sampled using  Monte-Carlo technique by choosing test particles (we use $N_{test}=100$ for each nucleon) with appropriate positions and momenta. As the the projectile and target nuclei propagate in time, the test particles move in a mean field and occasionally suffer two-body collisions, with probability determined by the nucleon-nucleon scattering cross section, provided the final state of the collision is not blocked by the Pauli principle. The mean-field propagation is done using the lattice Hamiltonian method which conserves energy and momentum very accurately \cite{Lenk}. The mean field potential is given by:
\begin{equation}
U(\rho)=A\bigg{\{}\frac{\rho}{\rho_0}\bigg{\}}+B\bigg{\{}\frac{\rho}{\rho_0}\bigg{\}}^{\sigma}+\frac{C}{\rho_0^{2/3}}\nabla_r^2\bigg{\{}\frac{\rho(\vec{r})}{\rho_0}\bigg{\}}
\label{Lenk_potential}
\end{equation}
where first two term represents zero range Skyrme interaction and the derivative term does not affect nuclear matter properties but in a finite system it produces quite realistic diffuse surfaces and liquid drop binding energies. This can be archived for $A=$-2230.0 MeV $fm^3, B$=2577.85 MeV $fm^{7/2}, \sigma=$7/6, $\rho_0=0.16$ and $c$=-6.5 MeV$fm^{5/2}$ \cite{Lenk}. Two body collisions are calculated as in Appendix B of ref. \cite{Dasgupta_BUU1}, except that pion channels are closed, as there will not be any pion production in this energy regime.\\
\indent
To explain clustering in heavy ion reaction, one needs an event-by-event computation in transport calculation. Bauer et. al. proposed the following method \cite{Bauer}. Due to collision between projectile nucleus of mass $A_p$ and target nucleus of mass $A_t$, for each event two body collisions are checked between $(A_p+A_t)N_{test}$ test particles. Test particle cross-sections are reduced to $\sigma_{nn}/N_{test}$; the collisions are further reduced by a factor $N_{test}$ but if a collision happens between two test particles $i$ and $j$ then not only these two change momenta but in addition $N_{test}-1$ test particles closest to $i$ in phase space suffer the same momentum change as $i$; also $N_{test}-1$ test particles closest to $j$ in phase space are given the same momentum change as $j$. Physically this corresponds to nucleons colliding. For conserving energy and momentum simultaneously, one can define $\langle\vec{p_i}\rangle=\frac{\sum\vec{p_{i}}}{N_{test}}$; similarly $<\vec{p_j}>$.  One then considers a collision between $<\vec{p_i}>$ and $<\vec{p_j}>$ and obtain a $\vec{\Delta p}$ for $<\vec{p_i}>$ and $-\vec{\Delta p}$ for $<\vec{p_j}>$.  This $\vec{\Delta p}$ is added to all $\vec{p_{i}}$'s and $-\vec{\Delta p}$ to all $\vec{p_{j}}$'s.  This conserves both energy and momentum as one progresses in time. For the second event new Monte-Carlo sampling of $A_p$ on $A_t$ will be started at time zero, similarly for event 3, event 4 etc. The calculation of the collision part becomes very time consuming and for this case within each time step two body collision is need to be checked between $(A_p+A_t)N_{test}$ test particles. Now, to study nuclear liquid gas phase transition one needs to simulate collisions between fairly large nuclei. Therefore, it is very difficult to handle this operation with the existing model. Hence one has to modify the transport model so that it can be used for fairly large nuclei.\\
\indent
To overcome this problem, the fluctuation added BUU method is modified in the following way. $N_{test}$ Monte-Carlo simulations of $A_p$ nucleons with positions and momenta and $N_{test}$ simulations of $A_t$ nucleons with positions and momenta is to be done as before . As in cascade calculation \cite{Dasgupta_BUU1,Mallik11,Mallik9} for nucleon-nucleon collisions 1 on 1'(event1), 2 on 2'(event2) etc are considered with cross-section $\sigma_{nn}$. For event 1, within each time step, $nn$ collisions only between 1 and 1' (i.e. between first $(A_p+A_t)$ test particles) will be considered. The collision is checked for Pauli blocking. If a collision between $i$ and $j$ in event 1 is allowed, ref. \cite{Bauer} is to be followed and $N_{test}-1$ test particles closest to $i$ are to be picked and the same momentum change $\Delta \vec{p}$ of them as ascribed to $i$ is to be given. Similarly $N_{test}-1$ test particles closest to $j$ are to be selected and these  are to be ascribed the momentum change $-\Delta \vec{p}$, the same as suffered by $j$. As a function of time this is continued till event 1 is over. For Vlasov propagation all test particles are utilised. For event 2 one has to return to time $t$=0, the original situation (or a new Monte-carlo sampling for the  original nuclei), follow the above procedure but consider $nn$ collisions only between 2 and 2' \{i.e. between $(A_p+A_t)+1$ to $2(A_p+A_t)$ test particles\}. This can be repeated for as many events as one needs to build up enough statistics. Finally to identify fragments, two test particles are considered as the part of the same cluster if the distance between them is less than or equal to $2$ fm \cite{Mallik16,Mallik22}.\\
\begin{figure}[!h]
\begin{center}
\includegraphics[width=0.9\columnwidth,keepaspectratio=true,clip]{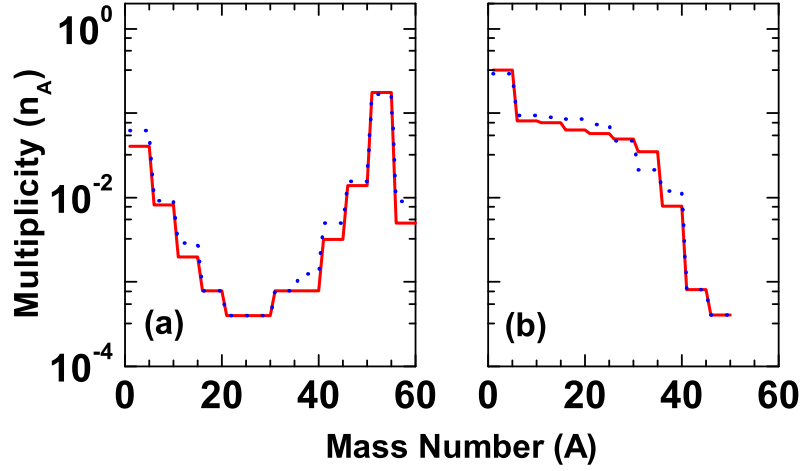}
\caption{Comparison of mass distribution  calculated according to the existing (blue dotted lines) and the modified (red solid lines) BUU prescription. The average value of 5 mass units are shown. The cases are for central collision of mass 40 on mass 40 for two different beam energies (a) 25 and (b) 50 MeV/nucleon calculated at $t=$200 $fm/c$.}
\label{Mass_distribution_two_methods_BUU}
\end{center}
\end{figure}
\indent
Before applying the modified method in phase transition study, at first one has to check whether modified prescription results are comparable with the existing BUU prescription or not. For this, simulations have to be done by using the both methods for a central collision reaction of symmetric system $^{40}Ca+^{40}Ca$. Fig. \ref{Mass_distribution_two_methods_BUU} shows the comparison of mass distribution obtained from existing and modified BUU prescription for beam energies 25 and 50 MeV/nucleon at $t=$200 $fm/c$. The results obtained from two methods are similar, because, (a) the number of collisions in an event are statistically the same. (b) In the original formulation the objects that collided were picked from a fine grain sampling of phase-space density. In the modified method these are picked from a coarse grain sampling of the same phase-space density. But many events are needed so statistically it should not matter. (c) Characteristics of scattering are the same. and
(d)The same Vlasov propagation is used.\\
\indent
The advantage of this method over the existing method is that here, for one event, $nn$ collisions need to be considered between $(A_p+A_t)$ test particles whereas in existing method, collisions need to be checked between $(A_p+A_t)N_{test}$ test particles. Hence, in the modified calculation, total number of combinations for two-body collision is reduced by a factor of 1/$N_{test}^2$. Since typically $N_{test}$ is of the order of 100 this is a huge saving in computation and has allowed us to treat reactions at different projectile energies and impact parameter described in the next sections. One bonus of this prescription is that one sees some common ground between the BUU approach and the "quantum molecular dynamics" approach \cite{Aichelin}.\\
\indent
At the end of the transport calculation i.e. at freeze-out stage, we get different clusters of finite number of test particles with known position and momenta. By knowing the number of test particles present in the cluster one can get the mass, and by knowing the position and momenta of these test particles one can calculate the potential and kinetic energies respectively. By adding kinetic and potential energy the excited state energy of the cluster can be obtained. However, to know excitation one needs to calculate the ground state state energy also. This is done by applying the Thomas Fermi method for a spherical (ground state) nucleus having mass equal to the cluster mass. Knowing PLF mass and its excitation, the freeze-out temperature as well as decay of excited clusters are calculated by using the canonical thermodynamic model CTM\cite{Dasgupta_CTM}.\\
\section{Identification of freeze-out}
Our first aim is to identify the freeze-out time when one can safely stop transport calculation and switch over to the statistical model. In order to do that we have investigated the time dependence of (i) mass of largest and second largest cluster and (ii) isotopy of momentum distribution in largest and second largest cluster from BUU calculation. Indeed in the binary collisions we consider the largest and second largest cluster are always the residues of projectile and target. The first signal can therefore help us to determine the time when the projectile and target are completely separated while the second one will point to the attainment of thermalization of these residues.\\
\begin{figure}[!h]
\begin{center}
\includegraphics[width=0.8\columnwidth,keepaspectratio=true,clip]{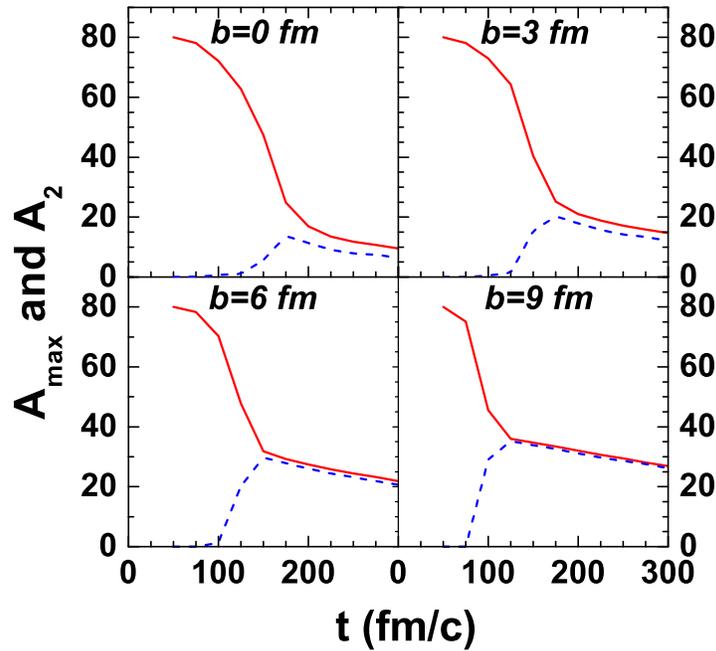}
\caption{Variation of average mass of largest cluster $A_{max}$ (red solid lines) and second largest cluster $A_2$ (blue dashed lines) with time as calculated from BUU model for (a) $b=$0 fm, (b) $b=$3 fm, (c) $b=$6 fm, (d) $b=$9 fm at projectile beam energy 100 MeV/nucleon.}
\end{center}
\end{figure}
\indent
The dependence of average size of the largest and the second largest cluster with time for symmetric system $^{40}Ca+^{40}Ca$ at projectile beam energy 100 MeV/nucleon but four different impact parameters ranging from central to peripheral collisions is displayed in Fig.2. The nature of variation is almost identical in each impact parameter. At $t$=50 fm/c, there was just one system comprising of both projectile and the target nuclei, hence the size of the largest cluster ($A_{max}$)is close to 80 and the second largest ($A_2$) is close to 0. With the progress of time, the size of the largest cluster decreases gradually as the system fragments as well as there is evaporation of light clusters and nucleons. For central collision, the size of the participant zone is maximum which results in faster disintegration therefore the rate of decrease largest cluster size is also maximum, while with the increase of impact parameter participant size decreases which gradually reduces the rate of decrease of largest cluster size. The size of the second largest starts from zero, gradually increases as the target and projectile crosses each other and reaches a maximum when they are completely separated and then again decreases because of secondary decay, and settles to  a final value.  The evolution of the largest and that of the second largest cluster is  pretty similar after the second largest cluster reaches its maximum and the evolution coincides for the most peripheral collisions.  This is only because we are dealing with identical size of projectile and target, and would change if one considers an asymmetric entrance channel.\\
\begin{figure}[!h]
\begin{center}
\includegraphics[width=0.8\columnwidth,keepaspectratio=true,clip]{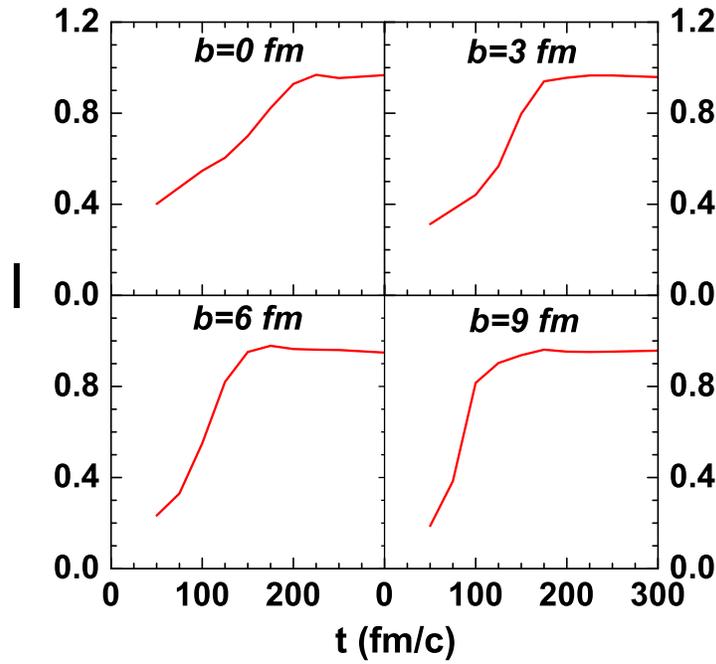}
\caption{Variation of isotropy of momentum distribution ($I$) of the largest cluster with time calculated from BUU model for (a) $b=$0 fm, (b) $b=$3 fm, (c) $b=$6 fm, (d) $b=$9 fm at projectile beam energy 100 MeV/nucleon.}
\end{center}
\end{figure}
\indent
The time evolution of the average isotropy of the momentum distribution ($I$) of the largest and second largest cluster is shown in Fig. 3. This observable indicates the thermalization of the system, and hence the ideal time to switch over from the dynamical model to statistical model. This is defined through the following equations. Let the beam direction is along $z$ axis and for a given event, out of total $(A_p+A_t)N_{test}$ test particles, only N test particles form a cluster i.e. the mass of the cluster is $N/N_{test}$. The average momentum of the cluster along $k=$ $x$, $y$ and $z$ direction can be calculated from the relation
\begin{equation}
P_k=\frac{1}{N}\sum_{i=1}^{N} {p_{k_i}}
\end{equation}
where $p_{k_i}$ is the $k$ component of momentum of the $i$-th test particle. The isotropy in momentum distribution can be defined as
\begin{equation}
I=\frac{\frac{1}{N}\sum_{i=1}^{N}(p_{x_i}-P_x)^2+\frac{1}{N}\sum_{i=1}^{N}(p_{y_i}-P_y)^2}{2\times\frac{1}{N}\sum_{i=1}^{N}(p_{z_i}-P_z)^2}
\end{equation}.
\begin{figure}[b]
\begin{center}
\includegraphics[width=0.8\columnwidth,keepaspectratio=true,clip]{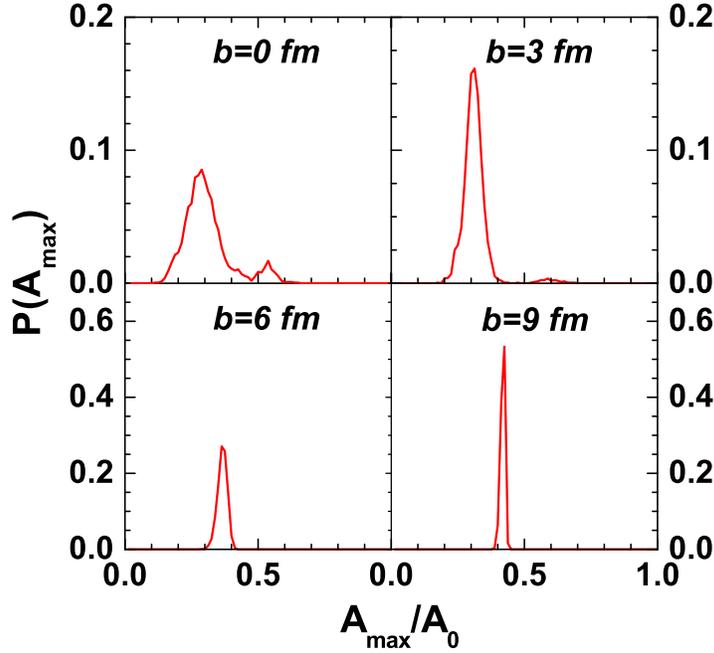}
\caption{Largest cluster probability distribution $P(A_{max})$ at freeze-out stage ($t$=175 $fm/c$) for constant projectile beam energy 100 MeV/nucleon but four different impact parameters (a )$b=$0 fm, (b) $b=$3 fm, (c) $b=$6 fm, (d) $b=$9 fm calculated from BUU model.}
\end{center}
\end{figure}
\indent
The quantity is defined such that it is less than 1 when the system is not fully thermalized and still there are some test particles having significant momentum in the beam direction. This will reduce the isotropy. Initially during the overlapping stage of the projectile and target nuclei, the isotropy is less than unity. With the increase of time it gradually increases and finally becomes unity when complete thermalization is achieved. Comparing Fig. 2 and 3 it can be concluded that $I\approx1$ is archived almost at the same time when the second largest cluster size is also maximum. This freeze-out time varies from about  $t_{f}=150$ fm/c for most peripheral collision to about $t_{f}=200$ fm/c for central collision. For simplicity, we have stopped  the dynamical calculation at $t=175$ fm/c for all impact parameters. Accounting for the precise impact parameter dependence of the freeze-out time would only marginally affect the distributions shown in this article, and would not affect any of our conclusions which are essentially based on the qualitative properties of the distributions. With changing the projectile beam energy, the freeze-out time will also change, for example by doing similar analysis for  $40$ MeV per nucleon reaction freeze-out time is determined as approximately $t=400$ fm/c.\\
\begin{figure}[b]
\begin{center}
\includegraphics[width=0.8\columnwidth,keepaspectratio=true,clip]{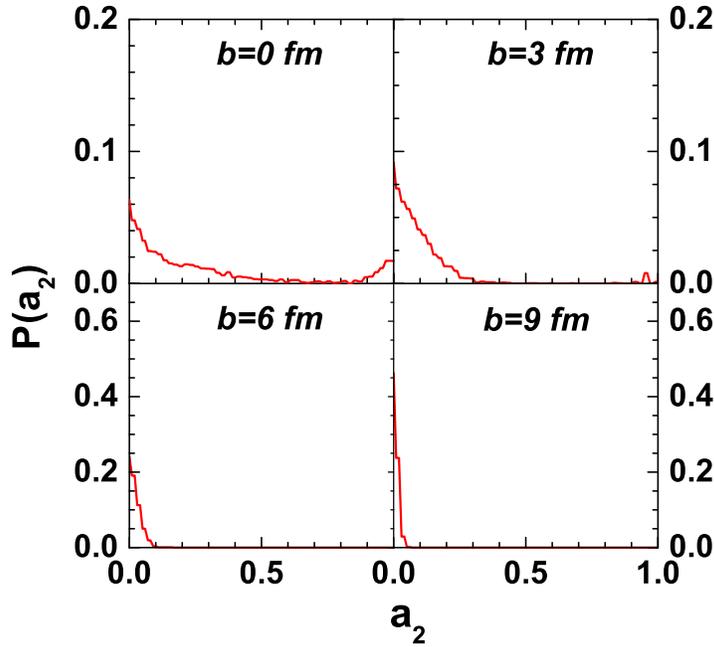}
\caption{Probability distribution of normalised mass asymmetry of two largest masses  $P(\mathrm{a}_2)$ at freeze-out stage ($t$=175 $fm/c$) for constant projectile beam energy 100 MeV/nucleon but four different impact parameters (a) $b=$0 fm, (b) $b=$3 fm, (c) $b=$6 fm, (d) $b=$9 fm calculated from BUU model.}
\end{center}
\end{figure}
\begin{figure}[t]
\begin{center}
\includegraphics[width=0.8\columnwidth,keepaspectratio=true,clip]{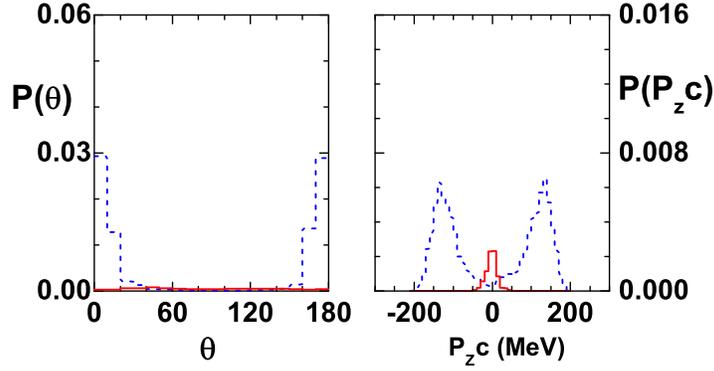}
\caption{Largest cluster scattering angle (left panel) and momentum (right panel) probability distribution for $A_{max} \geq 37$ (red lines) and $A_{max}<37$ (blue lines) for central collisions ($b=$0 fm) at projectile beam energy 100 MeV/nucleon calculated from BUU model at freeze-out time $t$=175 $fm/c$. The average value of 10 degrees and 10 MeV are shown for angle and momentum respectively.}
\end{center}
\end{figure}
\begin{figure}[t]
\begin{center}
\includegraphics[width=0.8\columnwidth,keepaspectratio=true,clip]{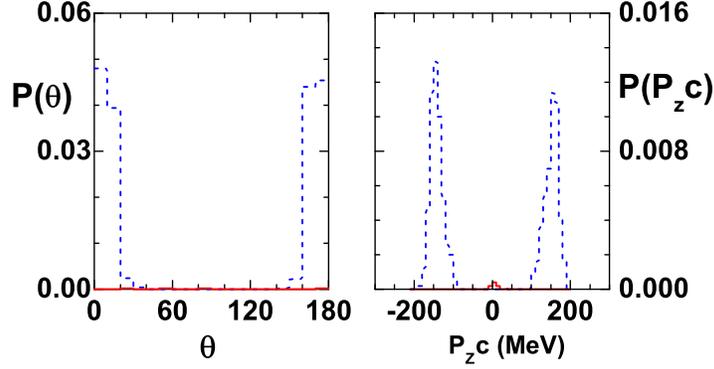}
\caption{Same as Fig. 6 except the impact parameter is 3 fm.}
\end{center}
\end{figure}
\section{Dynamical bimodality}
In this section we will concentrate on the behavior of probability distribution of largest cluster ($A_{max}$) and asymmetry of largest and second largest cluster ($\mathrm{a}_2=(A_{max}-A_2)/(A_{max}+A_2)$) calculated at the end of transport simulation at freeze-out condition. $P(A_{max})$ and $P(\mathrm{a}_2)$ distribution at constant projectile beam energy 100 MeV/nucleon but four different impact parameters ranging from central to peripheral collisions are shown in Fig. 4 and 5 respectively.\\
\begin{figure}[b]
\begin{center}
\includegraphics[width=0.8\columnwidth,keepaspectratio=true,clip]{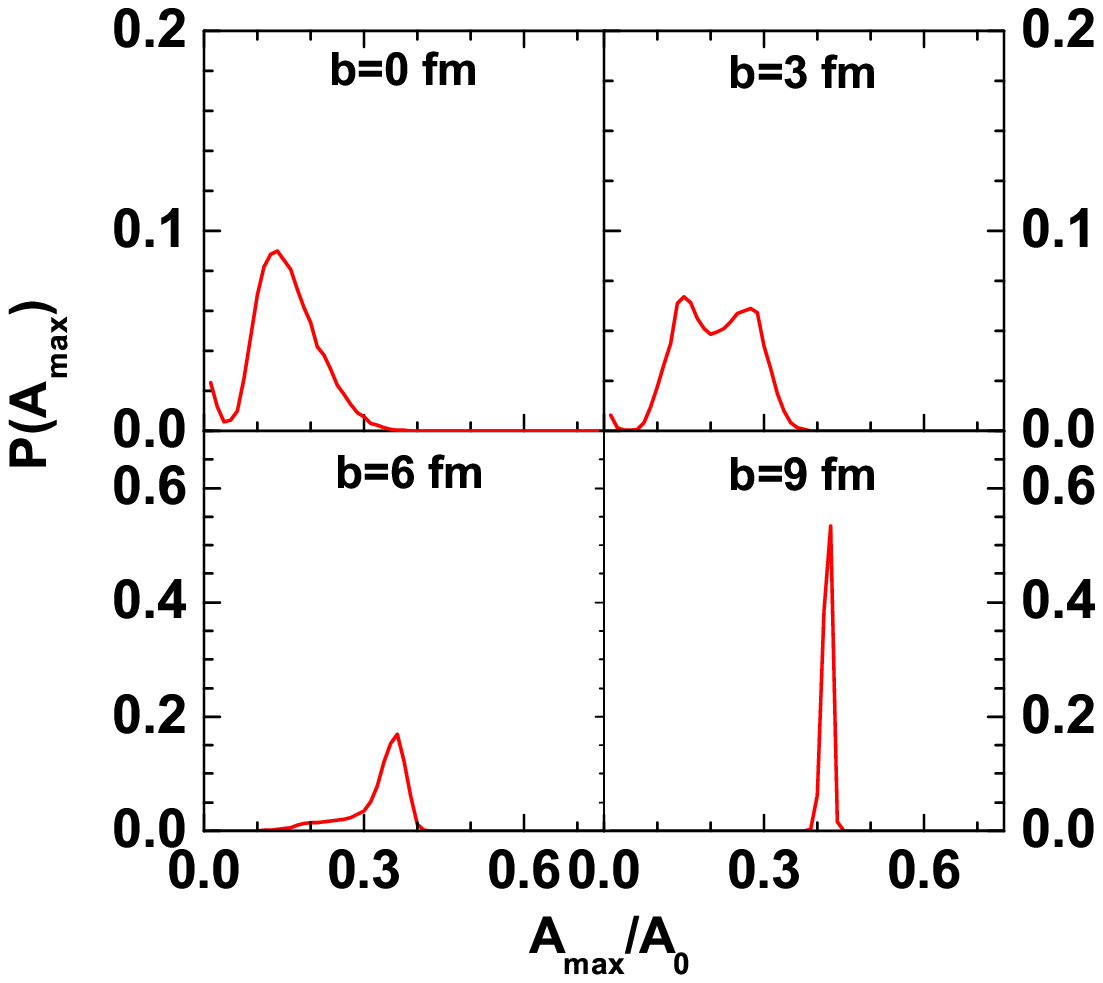}
\caption{Largest cluster probability distribution $P(A_{max})$ after econdary decay for constant projectile beam energy 100 MeV/nucleon but four different impact parameters (a) $b=$0 fm, (b) $b=$3 fm, (c) $b=$6 fm, (d) $b=$9 fm calculated from BUU model.}
\end{center}
\end{figure}
\indent
For central collision ($b$=0 $fm$), two peaks of $P(A_{max})$ as well as $P(\mathrm{a}_2)$ distributions are seen which can be interpreted as dynamical bimodality very similar to the phenomenon described in \cite{lefevre}. Fluctuations in the collision rates lead to fluctuations in the momentum distribution, that is in the degree of stopping of the reaction. We have fixed a mass cut of $A_{cut}=37$ (corresponds to the minimum between the two peaks at $b=$0 fm to distinguish the two event classes as it corresponds to the minimum between the two peaks. Fragments with $A_{max} \geq A_{cut}$ represent stopped events having nearly zero z-component (beam direction) of momentum and scattered isotropically in the centre of mass frame where as fragments with $A_{max}<A_{cut}$ represent crossed events having high z-component of momentum and scattered either in the forward direction (projectile like fragments) or backward direction (target like fragments). This is shown in Fig. 6. For non-central collisions, due to lesser overlapping of the projectile and target, almost all are crossed-events and only liquid phase is present. For $b$=3 fm This is shown in Fig. 7.\\
\begin{figure}[t]
\begin{center}
\includegraphics[width=0.8\columnwidth,keepaspectratio=true,clip]{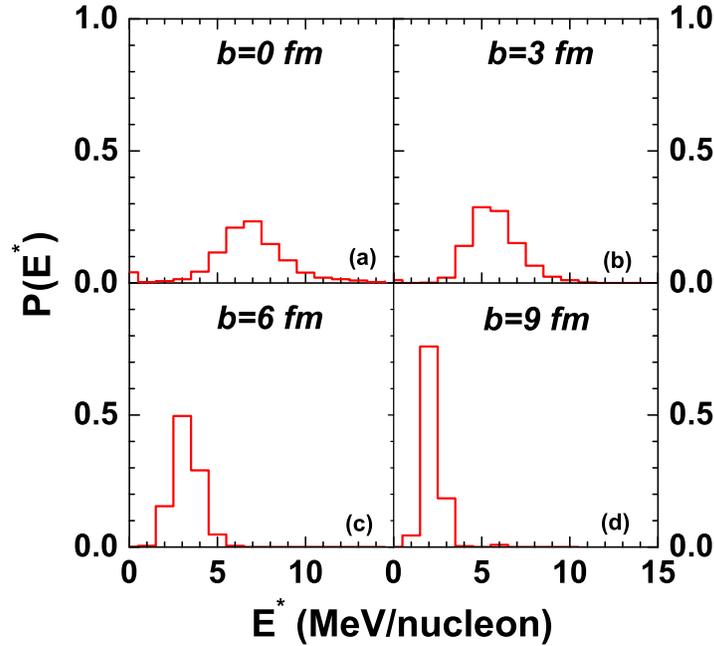}
\caption{Excitation ($E^*$) probability distribution for largest and second largest clusters at constant projectile beam energy 100 MeV/nucleon but four different impact parameters (a) $b=$0 fm, (b) $b=$3 fm, (c) $b=$6 fm, (d) $b=$9 fm.}
\end{center}
\end{figure}
\begin{figure}[t]
\begin{center}
\includegraphics[width=0.8\columnwidth,keepaspectratio=true,clip]{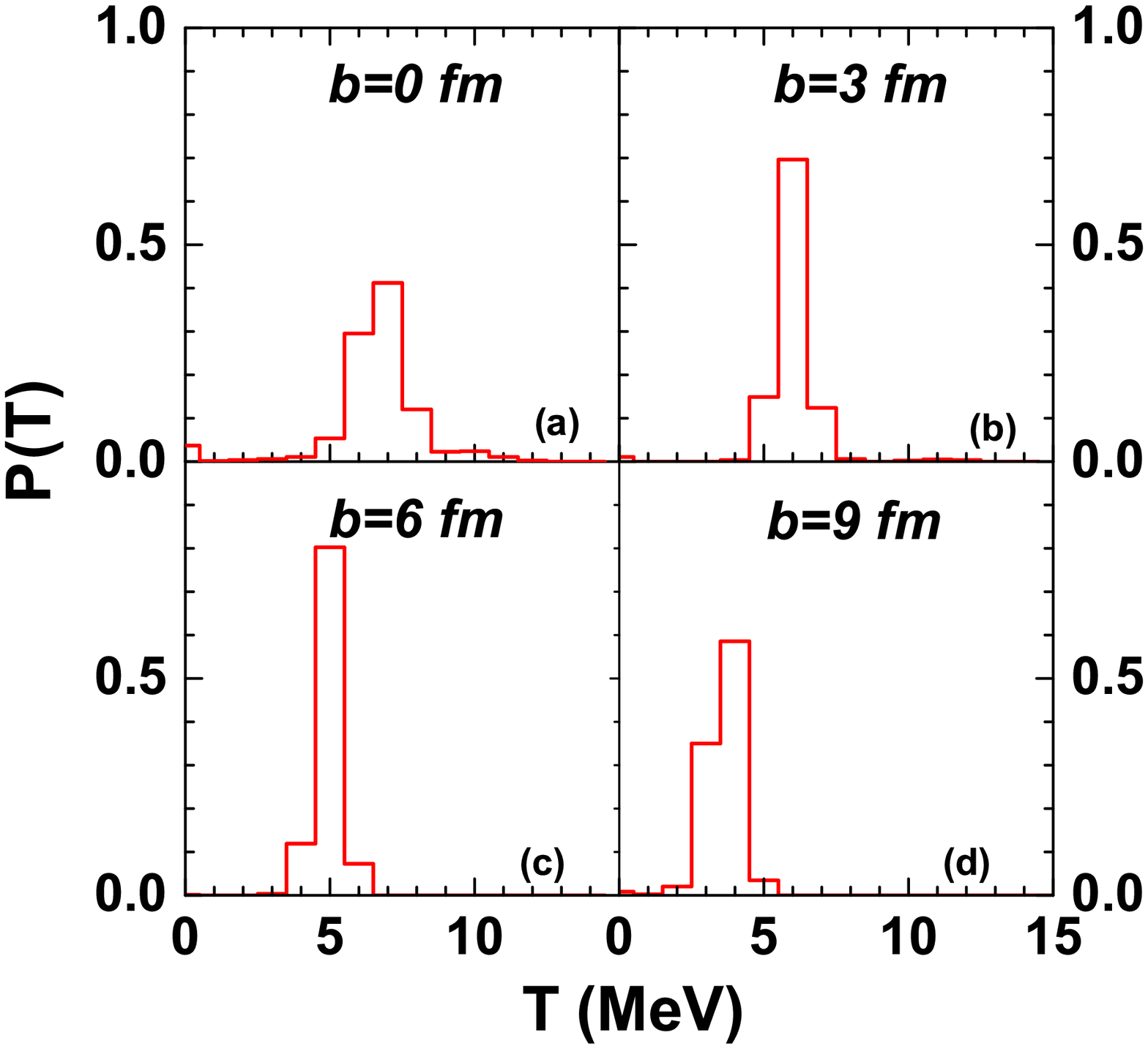}
\caption{Temperature ($T$) probability distribution for largest and second largest clusters at constant projectile beam energy 100 MeV/nucleon  but four different impact parameters (a) $b=$0 fm, (b) $b=$3 fm, (c) $b=$6 fm, (d) $b=$9 fm.}
\end{center}
\end{figure}
\begin{figure}[t]
\begin{center}
\includegraphics[width=0.9\columnwidth,keepaspectratio=true]{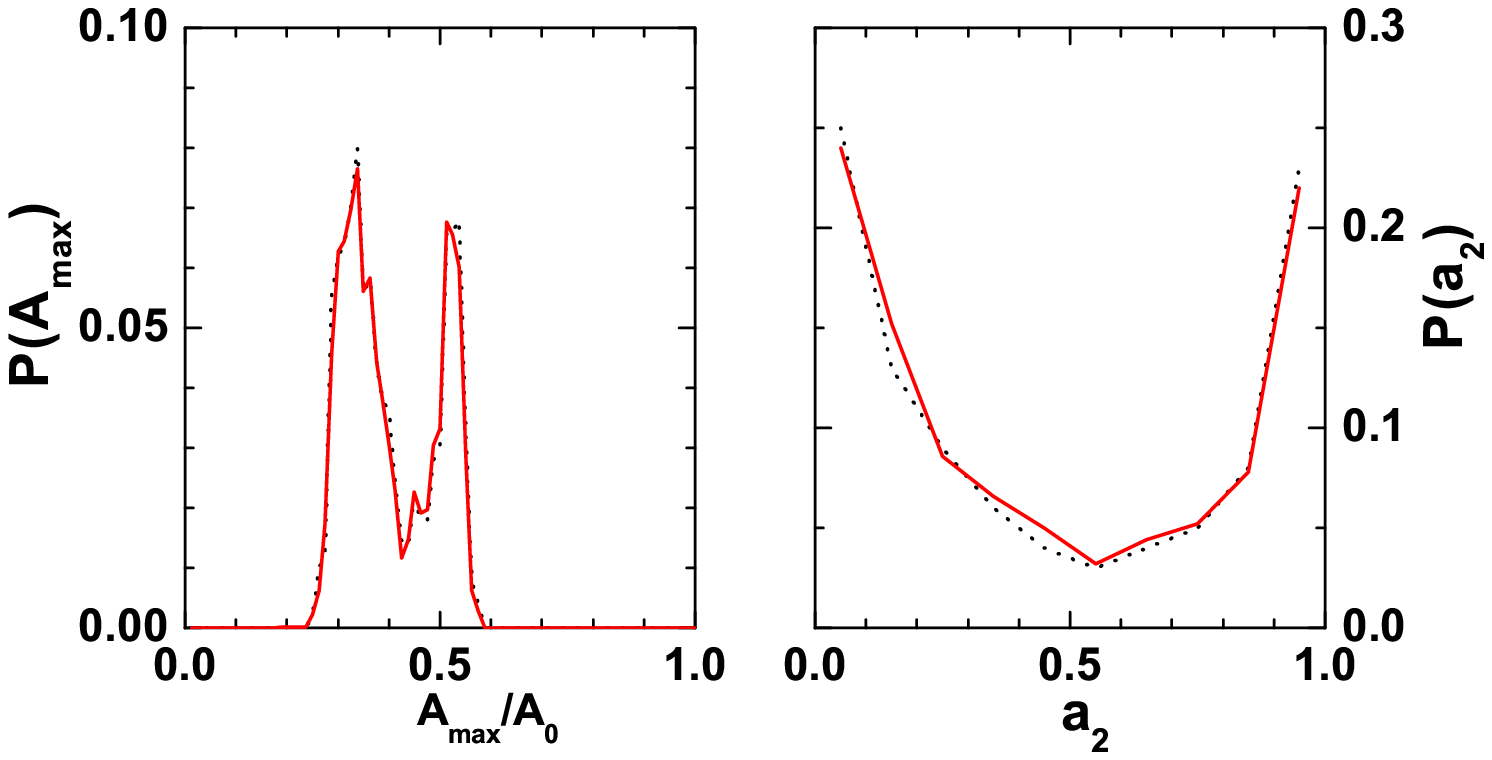}
\caption{Probability distribution largest cluster $P(A_{max})$ (left panel) and normalised mass asymmetry of two largest masses  $P(\mathrm{a}_2)$ (right panel) studied after BUU model calculation (black dashed line) and CTM calculation (red solid line) for central collisions ($b=$0 fm) at projectile beam energy 40 MeV/nucleon.}
\end{center}
\end{figure}
\section{Statistical bimodality}
The distribution plotted in Fig. 4 and 5 can be defined as freeze-out distribution and can still evolve in subsequent time because of secondary decay which have been calculated by switching over to the Canonical Thermodynamical model calculation from the transport one. In Fig. 8, we have plotted the probability distribution of the largest cluster for these four impact parameters. The ones at b=0 fm are structuresless and typical of multifragmentation reactions: the average excitation energy is so high in this case that both fully stopped and incompletely stopped events undergo multiple decay. As a consequence, the bimodality signal observed in Fig. 4 disappears. At mid-central collision, the situation is reversed. The probability distribution of the largest cluster now shows a bimodal behaviour (but not dynamical bimodality in Fig. 4) which is indicative of existence of two phases simultaneously.\\
\indent
The indication of thermal phase transition will be more clear if one concentrate on the excitation energy (Fig. 9) and temperature (Fig. 10) spectrum of the largest and second largest with varying centrality of the reaction. The step size selected for displaying these distributions are 1 MeV/A for the excitation energy and 1 MeV for the temperature. For the excitation energy in central collision, there is a small peak at low excitation that corresponds to the second largest cluster of the stopped event which is small in size and has less excitation. Using these excitation energies from the transport code as input to the statistical model code, temperature is obtained. At $b$=3 fm, the excitation spectrum is broadened but the temperate distribution is quite sharp indicating its connection to the thermal phase transition during which temperature remains constant.\\
\indent
The bimodal behavior described above, however strongly depends on the entrance channel conditions. In particular, central collisions at lower bombarding energy (40 MeV/nucleon) where the effect of secondary decay is less therefore the freeze-out distribution is not distorted by secondary decay and bimodal behaviour can be observed both after transport calculation, and after the statistical model calculation. This is shown in Fig. 11.\\
\section{Summary}
In order to study nuclear liquid gas from transport model, a simplified yet accurate method of Boltzmann-Uehling-Uhlenbeck (BUU) model is developed \cite{Mallik10} which allow calculation of fluctuations in systems much larger than what was considered feasible in a well-known and already existing model.\\
\indent
We have analyzed the bimodal behavior for the symmetric system $^{40}Ca+^{40}Ca$ with varying centrality of the reaction as well as bombarding energies,
as predicted by a two-step model. The entrance channel dynamics is described by the BUU transport equation, which is coupled to the statistical CTM decay model at the time of local equilibration of the primary fragments produced in the collision.\\
\indent
Based on combined theoretical simulation, it is observed that depending on the incident energy and impact parameter of the reaction, both entrance channel and exit channel effects can be at the origin of the experimentally observed bimodal behavior. Specifically, fluctuations in the reaction mechanism induced by fluctuations in the collision rate, as well as thermal bimodality directly linked to the nuclear liquid-gas phase transition are observed in simulations \cite{Mallik18}. These results indicate that heavy-ion reactions at intermediate energies can be used in the laboratory to study dynamical as well as statistical bimodalities

\end{document}